\documentclass[twocolumn,prb,aps,10pt,superscriptaddress]{revtex4-1}

\usepackage{amsfonts,amssymb}
\usepackage[sumlimits,intlimits]{amsmath}
\usepackage{graphics}
\usepackage{graphicx}
\usepackage{mathrsfs}
\usepackage{textcomp}
\usepackage{verbatim}
\usepackage{hyperref}    
\usepackage{bm}          

\pdfoutput=1

\newcommand{\be}{\begin{equation}}
\newcommand{\ee}{\end{equation}}
\newcommand{\lb}{l_B}

\newcommand{\ntot}{\nu_\text{tot}}
\newcommand{\Vt}{\widetilde{V}}
\newcommand{\RR}{\mathbf{R}}
\newcommand{\rr}{\mathbf{r}}
\newcommand{\nuc}{\nu_\text{c}}
\newcommand{\D}{\Delta_{1/3}}

\begin{document}

\title{The effect of electron dielectric response on the quantum capacitance of graphene in a strong magnetic field}

\author{Brian Skinner}
\affiliation{Fine Theoretical Physics Institute, University of Minnesota, Minneapolis, MN 55455, USA}
\author{G. L. Yu}
\author{A. V. Kretinin}
\author{A. K. Geim}
\author{K. S. Novoselov}
\affiliation{School of Physics and Astronomy, University of Manchester, Manchester, M13 9PL, United Kingdom}
\author{B. I. Shklovskii}
\affiliation{Fine Theoretical Physics Institute, University of Minnesota, Minneapolis, MN 55455, USA}

\date{\today}

\begin{abstract}

The quantum capacitance of graphene can be negative when the graphene is placed in a strong magnetic field, which is a clear experimental signature of positional correlations between electrons.  Here we show that the quantum capacitance of graphene is also strongly affected by its dielectric polarizability, which in a magnetic field is wave vector-dependent.  We study this effect both theoretically and experimentally.  We develop a theory and numerical procedure for accounting for the graphene dielectric response, and we present measurements of the quantum capacitance of high-quality graphene capacitors on boron nitride.  Theory and experiment are found to be in good agreement.

\end{abstract}

\maketitle

\section{Introduction} \label{sec:intro}

Capacitance measurements provide a powerful experimental tool for probing correlated and quantum behavior of electron gases.  In particular, capacitance measurements reveal the electron density of states, which is affected both by single-particle and many-body interaction effects.  Capacitance measurements are especially instructive for two-dimensional electron gases (2DEGs), where the global electron density can be modulated widely and reversibly by electrostatic gating, permitting one to map out the density of states as a function of the electron density.  The study of graphene, in particular, has seen tremendous recent advancements from this technique, with recent experiments using capacitance measurements to observe such phenomena as Fermi velocity renormalization \cite{Yu2013ipi}, fractional quantum Hall phase transitions \cite{Feldman2013fqh}, and Hofstadter's butterfly \cite{Hunt2013mdf}.

In the typical experimental setup, a graphene layer is separated by a distance $d$ from a parallel, metallic gate by an insulating spacer with dielectric constant $\kappa$.  A voltage source maintains a fixed (but tunable) electrochemical potential difference $V$ between the graphene and the gate electrode.  The resulting differential capacitance per unit area, $C = e dn/dV$, where $e$ is the two-dimensional (2D) electron charge and $n$ is the electron density, can be written generically as
\be
C = \left( C_g^{-1} + \frac{1}{e^2} \frac{d\mu}{dn} \right)^{-1}. \label{eq:Cdef}
\ee
Here, $C_g = \kappa /4 \pi d$ is the standard ``geometric" capacitance of a plane capacitor with thickness $d$ (in Gaussian units) and $\mu$ is the electron chemical potential, which we define relative to the Dirac point.  The quantity $dn/d\mu$ is the thermodynamic density of states (TDOS), and $e^2 dn/d\mu$ is commonly referred to as the ``quantum capacitance," which effectively adds in series with the normal, geometric value.  For the purposes of this paper, it is convenient to define the effective capacitor thickness $d^* = \kappa/4\pi C$, so that Eq.\ (\ref{eq:Cdef}) can be written as $d^* = d + d_Q$, where $d_Q$ is a ``quantum capacitance length" given by
\be 
d_Q = \frac{\kappa}{4 \pi e^2} \frac{d \mu}{d n} = d \left( \frac{C_g}{C} - 1 \right).
\label{eq:dQdef}
\ee 
This length is related to the Debye screening radius \cite{Skinner2010alc} $r_D$ by $d_Q = r_D/2$; for the purposes of this paper, $d_Q$ can be thought of as a renormalization of the capacitor thickness arising from the finite density of states.  When the TDOS is positive, the capacitance is reduced and the effective thickness $d^*$ of the capacitor is larger than the physical thickness $d$.

On the other hand, in the strongly-interacting regime a 2DEG can have \emph{negative} TDOS \cite{Bello1981dol, Luryi1988qcd, Kravchenko1990eio, Efros1990dos, Eisenstein1992nco, Eisenstein1994cot, Shapira1996toc, Dultz2000tso, Ilani2000ubo, Allison2006tdo, Shklovskii1986soo, Efros1992hai, Efros1992tdo, Pikus1993doe, Shi2002dsa, Fogler2004nsa, Efros2008ndo, Kopp2009coc, Li2011vlc, Tinkl2012lne}, which arises as a consequence of strong positional correlations between electrons.  Such negative TDOS implies a negative $d_Q$, and therefore a capacitance that is enhanced above the geometric value, as was first measured experimentally over two decades ago in Si MOSFETS and GaAs heterostructures\cite{Kravchenko1990eio, Eisenstein1992nco, Eisenstein1994cot, Shapira1996toc, Dultz2000tso, Ilani2000ubo}.  Generally speaking, the strong positional correlations that give rise to this ``negative compressibility" arise when the electron gas has a large ratio of interaction energy to kinetic energy.  For electron gases with parabolic dispersion (as in semiconductor quantum wells), this corresponds to a large value of the parameter $r_s = 1/\sqrt{\pi n a_B^2}$, where $a_B$ is the effective Bohr radius.

In graphene, however, the linear dispersion relation implies a ratio of Coulomb to kinetic energy that is independent of the electron density and is characterized by the effective fine structure constant
\be 
\alpha = \frac{e^2}{\kappa \hbar v} \approx \frac{2.2}{\kappa}.
\ee 
It is therefore not possible to reach the strongly-interacting regime just by reducing the electron density.  Instead, a strongly-correlated phase can be reached by applying a strong perpendicular magnetic field $B$, which effectively quenches the electron kinetic energy by Landau quantization.  In such a magnetic field the strongly-correlated regime corresponds \cite{Lozovik1975ctd} to small $n \lb^2$, where $\lb = \sqrt{\hbar c / eB}$ is the magnetic length ($\hbar$ is the reduced Planck constant and $c$ is the speed of light), or in other words to small overall filling factor $\ntot = 2 \pi n \lb^2$.

In a recent work \cite{Yu2013ipi}, Yu \textit{et. al.} (including three of the present authors) studied experimentally the quantum capacitance of graphene in a strong magnetic field, where the lowest ($N = 0$) Landau level (LL) is split into four nondegenerate sublevels by the exchange interaction.  It was shown that at such large magnetic fields the quantum capacitance length $d_Q$ is indeed negative in the middle of each of these lowest Landau level sublevels (LLLSs).  
Below we show that further improvements in the quality of our samples result in significantly lower disorder, which allows us to resolve additional features in the quantum capacitance at large field that can be explained quantitatively with a zero-disorder theory.

In a separate theoretical work \cite{Skinner2013gcp}, two of us studied the dependence of the quantum capacitance on the capacitor thickness $d$ and the filling factor $\nu$ of one of the LLLSs.  Our focus in Ref.\ \onlinecite{Skinner2013gcp} was largely on the case where the capacitor is thin enough that $d/\lb \lesssim 1$.  In this case the screening of electron interactions via image charges in the metal gate becomes important \cite{Skinner2010alc, Skinner2010svm}, and at $d/\lb \rightarrow 0$ the quantum capacitance length $d_Q$ approaches $-d$, so that the capacitance becomes greatly enhanced above the geometric value.  (For low-frequency capacitance measurements, $d_Q < -d$ is not possible, since it would imply a negative capacitance, which is forbidden by thermodynamic stability arguments \cite{Landau1984eoc}.)  

In this paper we focus our attention primarily on the case of $d/\lb > 1$, where the image charge screening effect is relatively weak and $|d_Q|$ is small compared to $d$, and we consider an effect that was largely unexplored in previous works: the screening of electron interactions by the dielectric response of the graphene itself.  We develop a simple theory to describe this effect and support our theory with experimental measurements of $d_Q$ in clean, gated graphene on boron nitride.  Our primary message in this paper is that, in the case $d/\lb > 1$, quantum capacitance measurements in graphene cannot be described quantitatively without accounting for this dielectric response.

Unlike in conventional semiconductor 2DEGs, in graphene the in-plane dielectric response is significant even in the limit of vanishing carrier density.  This robust response arises as a consequence of graphene's gapless spectrum, which implies an easily polarizable ``Dirac sea."  Generally speaking, the dielectric response tends to work against Coulomb-driven quantum capacitance effects, since it weakens the interaction between electrons by effectively spreading part of their charge across the plane of the 2DEG.   The strength of this response is described by the zero-frequency dielectric function $\epsilon(q)$, where $q$ is the wave vector.  In the absence of a magnetic field, $\epsilon(q)$ acquires a constant value \cite{Goerbig2011epg} $\epsilon = 1 + \pi \alpha/2$.  In the presence of a strong quantizing field, on the other hand, $\epsilon(q) - 1$ vanishes at small $q \lb$ due to the finite energy gap between adjacent LLs, so that the interaction between electrons at long distances is unscreened.  
A quantitative description of the effect of $\epsilon(q)$ on the quantum capacitance of graphene is the primary aim of this paper.  

The remainder of this paper is organized as follows.  In Sec.\ \ref{sec:theory} we present our theoretical method for calculating the capacitance including the effect of the dielectric response, and we present general formulas for $d_Q$ as a function of $\nu$ in the LLLS.  Sec.\ \ref{sec:experiment} briefly describes our devices and experimental setup and presents our raw data.  In Sec.\ \ref{sec:compare} we analyze our experimental results and show that they compare well with theory.  Sec.\ \ref{sec:discussion} discusses the implications of our results for devices made from double-layer graphene and for devices that are thin enough that image charge effects become important.  We conclude in Sec.\ \ref{sec:conclusion} with a summary and some further discussion.

\section{General theory and numerical procedure}
\label{sec:theory}

If one ignores the possible effect of image charges, as mentioned in the Introduction, then the TDOS of the 2DEG in graphene is identical to that of an electron gas with a coplanar neutralizing background.  (In reality, this neutralizing background is displaced by a distance $d$ from the plane of the 2DEG, and this is what gives rise to the constant geometric capacitance $C_g$ that adds in series with the ``quantum" part.)  If the energy per electron of this 2DEG is $E(\nu)$, then Eq.\ (\ref{eq:dQdef}) can be written
\be 
d_Q = \frac{\lb}{2} \frac{d^2}{d \nu^2} \left[ \frac{ \nu E(\nu)}{e^2/\kappa \lb} \right].
\label{eq:dQnu}
\ee
Thus, an accurate calculation of $E(\nu)$ provides an estimate of the quantum capacitance.  Throughout this paper we focus on a model that neglects finite temperature.

Generally speaking, $E(\nu)$ in the LLLS can be written in the form of a power-law expansion that obeys the requisite electron-hole symmetry of the LLLS.  In particular, Fano and Ortolani (FO) proposed the formula\cite{Fano1988ife}
\be 
\nu E(\nu) = E(1) \nu^2 + \frac{e^2}{\kappa \lb} \sum_{k = 3}^\infty a_k \left[ \nu (1 - \nu) \right] ^{k/2},
\label{eq:FO}
\ee 
where $a_k$ are numerical coefficients.  The first three coefficients $a_3$, $a_4$, and $a_5$ were estimated by FO\cite{Fano1988ife} for a 2DEG without any dielectric response.
In particular, the coefficient $a_3 = -0.782$ can be found by calculating the energy of a classical Wigner crystal with density $n = \nu/2\pi\lb^2$, which gives the leading order contribution to the energy at vanishingly small $\nu$.  The other coefficients were originally determined in Ref.\ \onlinecite{Fano1988ife} by fitting Eq.\ (\ref{eq:FO}) to Monte Carlo calculations of $E(\nu)$ at different values of $\nu$.

In this section we modify the FO formula to include the effects of the dielectric response of the graphene, $\epsilon(q)$.  In other words, we calculate revised values of $E(1)$ and the coefficients $a_k$ that properly account for the graphene's dielectric polarizability.  Following Ref.\ \onlinecite{Skinner2013gcp}, our approach to this calculation is to first calculate $E(1)$ using known properties of the $\nu = 1$ state, then calculate $E(\nu)$ at small $\nu$ by treating the system as a Wigner cyrstal.  Finally, we fit the resulting energies at small $\nu$ to the form of Eq.\ (\ref{eq:FO}) to find the coefficients $a_k$.  Once these coefficients are known, one can insert Eq.\ (\ref{eq:FO}) into Eq.\ (\ref{eq:dQnu}) to get $d_Q$, and therefore the total capacitance.

It should be noted that the FO formula [Eq.\ (\ref{eq:FO})] cannot capture the cusps in energy that are associated with fractional quantum Hall (FQH) states, so that Eq.\ (\ref{eq:FO}) should be thought of only as a ``backbone" energy for the LLLS that is punctuated by short, sharp cusps located at FQH fractions.  These cusps produce ``bumps" in the quantum capacitance at such fractions that are not captured by the FO approach \cite{Eisenstein1992nco, Eisenstein1994cot}, and which will be discussed further in the following section.

The dielectric function $\epsilon(q)$ that we use throughout this paper is that of pristine graphene, which was calculated for the LLLS within the random phase approximation in Ref.\ \onlinecite{Shizuya2007era}.  The full analytic form for $\epsilon(q)$ is not reproduced here, but within the LLLS it can be viewed as a smooth crossover between its small- and large-$q$ asymptotic limits:
\begin{align}
&\epsilon(q) \simeq 1 + 1.74 \alpha q \lb , & & q \lb \ll 1 \\
&\epsilon(q) \simeq 1 + \pi \alpha/2, & & q \lb \gg 1.
\label{eq:epsconst}
\end{align}
We note that within the LLLS, the dielectric function is independent of $\nu$.  This independence can be viewed as a consequence of the symmetry of positive and negative LL energies about the $N = 0$ LL \cite{Shizuya2007era}.  In principle, electron interactions provide additional corrections to $\epsilon(q)$ that go beyond the random phase approximation.  These have been considered in Ref.\ \onlinecite{Sodemann2012ict}, but are relatively small and are not considered here.

The energy of the filled LLLS, $E(1)$, can be calculated using the general expression for the energy $E(\nu)$:
\be 
E(\nu) = \frac{n}{2} \int d^2r V(r) \left[ g(r) - 1 \right].
\label{eq:Eg}
\ee
Here, $V(r)$ is the electron-electron interaction law, $g(r)$ is the electron pair distribution function, and the $-1$ in the brackets comes from the interaction of the electrons with the uniform background.  For $\nu = 1$, where electrons occupy the $\nu = 1$ Laughlin liquid state, the pair distribution function $g(r) = g_1(r)$ is known\cite{Laughlin1990qhe}:
\be 
g_1(r) = 1 - \exp[-r^2/2 \lb^2].
\ee 
Inserting this expression for $g(r)$ into Eq.\ (\ref{eq:Eg}), setting $n = 1/(2 \pi \lb^2)$, and writing $V(r)$ in terms of its Fourier transform $\Vt(q) = 2 \pi e^2/[\kappa \epsilon(q) q]$ gives
\be 
E(1) = - \frac12 \frac{e^2}{\kappa \lb} \int_0^\infty \frac{\exp[-q^2 \lb^2/2]}{\epsilon(q)} \lb dq.
\label{eq:E1}
\ee
This integral can be done numerically for a given value of $\alpha$.  In Appendix \ref{app:FO} we give an approximate formula for its value at arbitrary values of $\alpha$.

Given the energy $E(1)$, one can arrive at values for the coefficients $a_k$ in the FO expression by calculating $E(\nu)$ over some finite range of $\nu$ and then performing a polynomial regression.  In our case, we use the range corresponding to small filling factors, $0 < \nu < \nuc$, where $\nuc \ll 1$, at which positional correlations are strong and one can closely approximate the energy $E(\nu)$ by calculating the energy of the Wigner crystal state.  The calculations presented here use $\nuc = 0.2$, which corresponds approximately to the liquid-solid transition point in the unscreened 2DEG\cite{Lam1984lta}.  
We verified, however, that our calculations are not substantially changed if $\nuc$ is made as small as $0.1$.

As mentioned above, at the small filling factors $\nu < \nuc$, the energy per electron is closely approximated by the energy $E_{WC}(\nu)$ of the Wigner crystal state.  We calculate this energy using a semiclassical (Hartree) approximation.  This calculation is straightforward, and is presented in Appendix \ref{app:hartree}.

Finally, we arrive at estimates for the coefficients $a_3$, $a_4$, and $a_5$ in Eq.\ (\ref{eq:FO}) by setting $E(\nu) = E_{WC}(\nu)$ for $0 < \nu < \nu_c$ and making a second-order polynomial fit of the quantity $[\nu E(\nu) - E(1) \nu^2]/[\nu (1 - \nu)]^{3/2}$ against $\sqrt{\nu (1 - \nu)}$.  Our result for these coefficients is parameterized as a function of $\alpha$ in Appendix \ref{app:FO}.  We note here only that our fitting produces a value of $a_3$ that is within $5\%$ of the value originally used by FO, $a_3 = -0.782$.  This is as expected, since the leading-order term of the expansion comes from the energy of a classical Wigner crystal with vanishingly small density, where the dielectric response plays no role (all relevant wavevectors $q$ satisfy $q \lb \ll 1$). 

As an additional check of our result, we verified that in the limit $\alpha \rightarrow 0$, where $\epsilon(q) = 1$ uniformly, our result for $E(\nu)$ is identical to the one originally proposed by FO\cite{Fano1988ife} to within $3\%$ at all $\nu$.  We also checked that our result reproduces the energy of the FQH states at $\nu = 1/5$ and $\nu = 1/3$ to within a few percent at all values of $\alpha$.  The calculation of these energies is presented in Appendix \ref{app:fqh}.

Shown in Fig.\ \ref{fig:newFO} is an example of our result for $E(\nu)$, calculated at $\alpha = 0.68$, which corresponds to the experimental data for graphene on boron nitride discussed in the following section.  As one can see, the graphene dielectric response provides a large reduction in the energy (in absolute value) at $\nu$ close to $1$, where the dominant contribution to the energy comes from interactions at distances $r \sim \lb$ (or $q \sim 1/\lb$).  In the limit of small $\nu$, on the other hand, the energy is largely unchanged from its unscreened value.  As expected, the energies of the $\nu = 1/5$ and $\nu = 1/3$ states, which are calculated independently of the fitting, lie very close to the $E(\nu)$ curve (and slightly below it).  

\begin{figure}[htb]
\centering
\includegraphics[width=0.48 \textwidth]{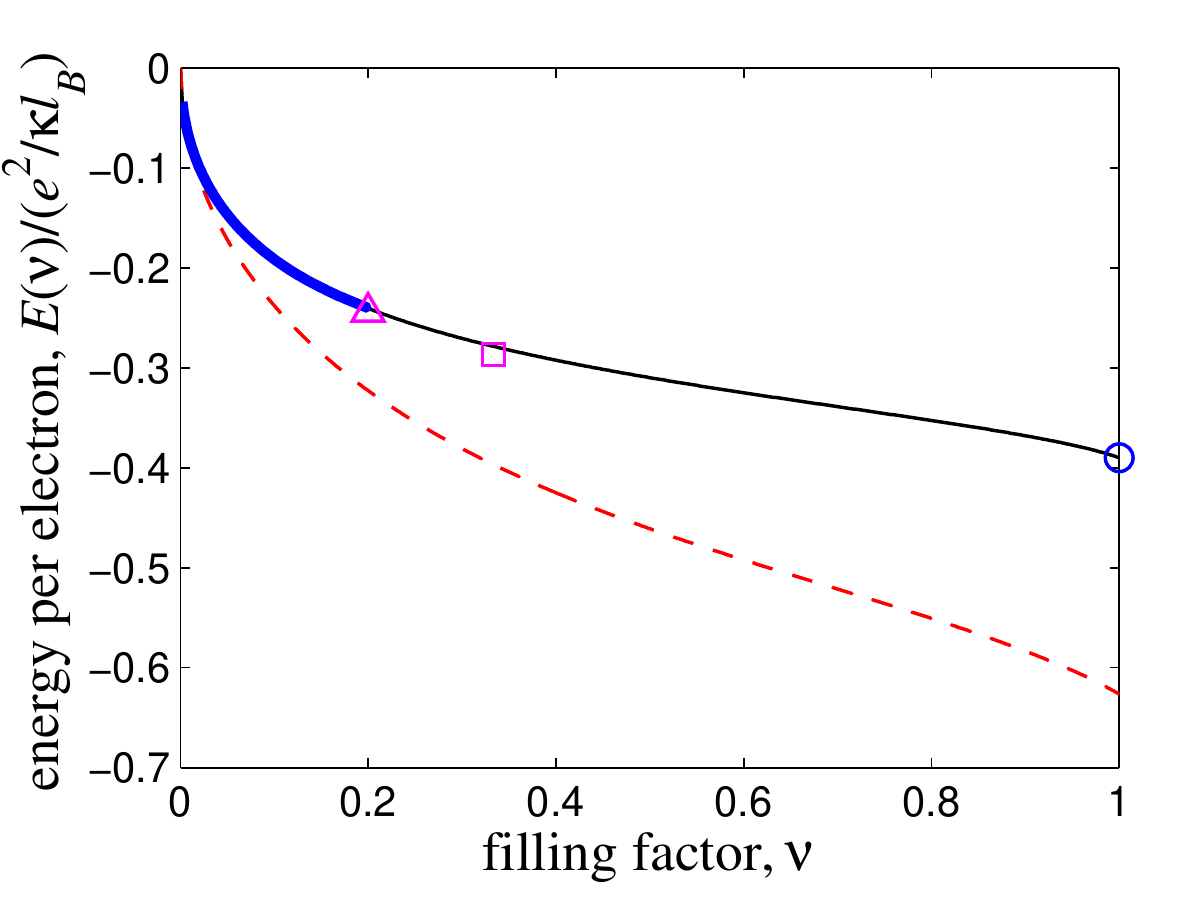}
\caption{(Color online) The energy per electron of a 2DEG in monolayer graphene in the LLLS at $\alpha = 0.68$.  The thin (black) line includes the effect of the graphene dielectric response, while the dashed (red) line uses $\epsilon(q) = 1$.  The thick (blue) line shown for $0 < \nu < 0.2$ is the result of the calculation of the Hartree energy of the WC state.  The (blue) circle at $\nu = 1$ is the calculation of Eq.\ (\ref{eq:E1}).  The (magenta) triangle and square show calculations of the energy of the $\nu = 1/5$ and $\nu = 1/3$ FQH state energies, which are calculated independently of the fitting and are used as checks of the interpolation formula.}
\label{fig:newFO}
\end{figure}

Using our results for $E(\nu)$ one can calculate $d_Q$ (and therefore the quantum capacitance) at arbitrary values of $\alpha$ and $\nu$ using Eqs.\ (\ref{eq:dQdef}) and (\ref{eq:FO}).  In principle, these equations can be combined into one algebraic formula, although it is too cumbersome to reproduce here.

Below we focus largely on the value of $d_Q$ at $\nu = 1/2$ and its dependence on magnetic field.  Using the procedure outlined in this section gives for the value of $\alpha$ mentioned above:
\be 
d_Q(\nu = 1/2) \approx -0.10 \lb, \hspace{5mm} (\alpha = 0.68).
\label{eq:dQ12}
\ee 
As we show below, Eq.\ (\ref{eq:dQ12}) provides a good description of our measured values of $d_Q$ at sufficiently large magnetic field.

Of course, our calculations in this section have ignored the possible effects of disorder, which at sufficiently small electron concentration overwhelm the electron-electron interactions and produce a rapid drop in the capacitance \cite{Shi2002dsa, Fogler2004nsa, Skinner2013gcp, Eisenstein1994cot}.   While the disorder-dominated regime is not a major focus of this paper, in the following section we provide some discussion of disorder effects in the context of experimental measurements.

\section{Experimental setup and raw data}
\label{sec:experiment}

\begin{figure}[htb]
\centering
\includegraphics[width=0.46 \textwidth]{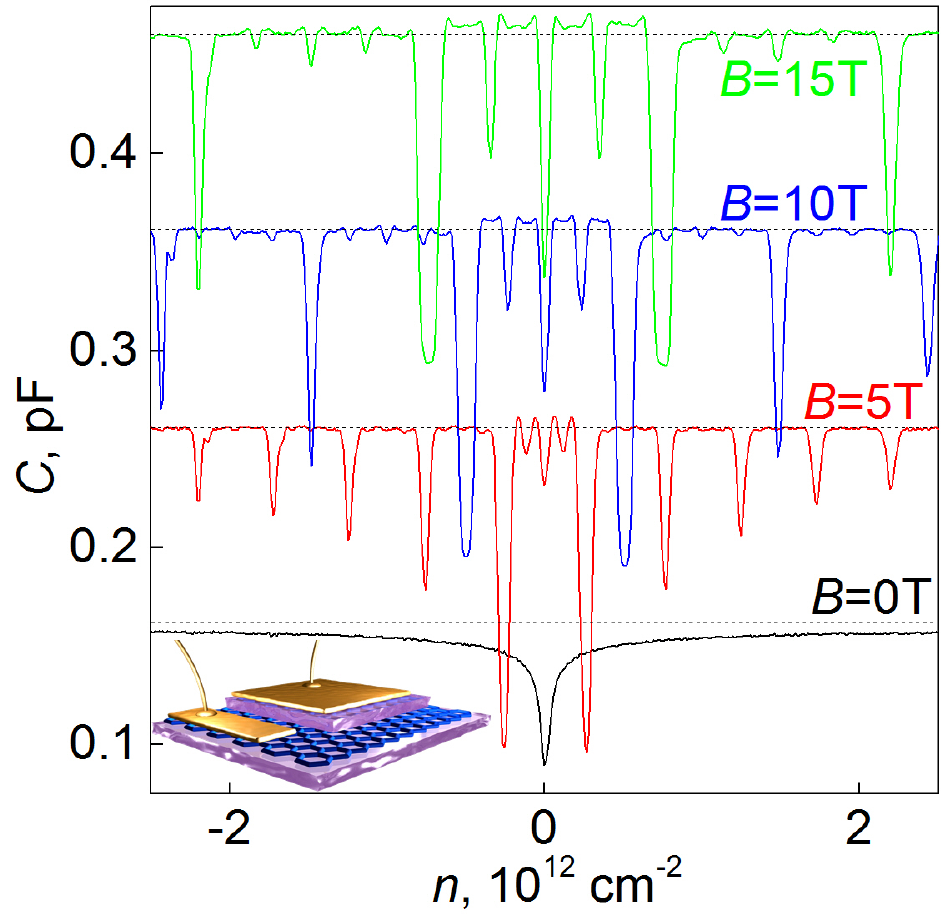}
\caption{(Color online) Differential capacitance as a function of electron density for one of our samples at $B=0$\,T (black curve), $5$\,T (red), $10$\,T (blue) and $15$\,T (green). Each curve at finite magnetic field is offset by $0.1$\,pF from the one below it. The magnitude of the geometrical capacitance for each curve is marked by a dashed line. 
The inset shows a schematic of the device.}
\label{fig:CV}
\end{figure}

In order to measure $d_Q$ experimentally, we constructed thin, gated graphene devices on boron nitride.  
Our devices consisted of a bottom graphene electrode separated from a Cr ($5$\,nm)/Au($50$\,nm) electrode by a thin dielectric layer -- typically $20$-$30$\,nm -- of hexagonal Boron Nitride (hBN). The whole sandwich rests on a thick layer of hBN (typically $50$\,nm) placed on a quartz substrate (see the inset of Fig.\ \ref{fig:CV} for a schematic of the device). Special care was undertaken in order to utilise only the flat and clean areas of the graphene/hBN sandwich for our capacitors, avoiding bubbles and contamination. Such selectivity allowed for a significant increase in the  homogeneity of our devices. For more information on device fabrication see Ref.\ \onlinecite{Yu2013ipi}.

The differential capacitance was measured by a capacitance bridge at the frequencies $1$-$20$\,kHz. The excitation voltage was in the range $1$-$20$\,mV, and was carefully chosen for each device in such a way that the modulation of the chemical potential doesn't exceed the broadening of the LL energies by inhomogeneities.  Measurements were taken at a temperature of $2$\,K over the range of magnetic fields $0 \leq B \leq 17.75$\,T.

Below we present results corresponding to one particular device, chosen for its low apparent disorder.  The thickness of this device, as measured by atomic force microscopy, was $d = 27.3$\,nm, so that at all $B > 1$\,T we indeed deal with the situation $d/\lb > 1$.  Our raw capacitance data for this device is presented in Fig.\ \ref{fig:CV}.  A small parasitic capacitance with magnitude $\sim 41$\,fF, arising from the wiring, has been subtracted from the data.  The capacitor area was $155$\,$\mu$m$^2$.

The effective dielectric constant of the capacitor was determined by looking at the voltage periodicity $\Delta V$ of the deep minima in $C$ at finite field, assuming these minima correspond to filled LLs ($\ntot = \pm 2, \pm 6, \pm 10$, etc.).  From this periodicity one can estimate $d/\kappa$ by equating $C_g = \kappa/4 \pi d$ with $e dn/dV \simeq e(4/2 \pi \lb^2)/\Delta V$.  This process gives $d/\kappa \approx 8.4$\,nm, so that $\kappa \approx 3.3$, which is consistent with known values of the dielectric constant of hBN.  The corresponding value of $\alpha$ is $\alpha \approx 0.68$.  The values of the geometric and parasitic capacitances were determined by fitting the $B = 0$ data to its known analytical form\cite{Ponomarenko2010dos, Yu2013ipi}, which is discussed in the following section.

\section{Analysis of experimental results}
\label{sec:compare}

In the previous sections we explained our procedures for calculating the quantum capacitance theoretically and for measuring it experimentally.  
In this section we discuss our measured results and compare them with the theory.  

In the absence of a magnetic field, $d_Q$ is positive and given by
\be 
d_Q = \frac{|n|^{-1/2}}{8 \alpha \sqrt{\pi}},
\label{eq:dQ0}
\ee 
as dictated by the finite (positive) TDOS of graphene resulting from the linear spectrum \cite{Ponomarenko2010dos}.  
From our data at $B = 0$, we find that $d_Q$ indeed remains linear in $|n|^{-1/2}$ for all electron densities $|n| \gtrsim 5\times 10^{10}$ cm$^{-2}$, with a slope that is consistent with our above estimate $\alpha = 0.68$.
At smaller $|n|$, $d_Q$ saturates, presumably due to the formation of disorder-induced electron/hole puddles \cite{Ponomarenko2010dos}.

\begin{figure}[htb]
\centering
\includegraphics[width=0.5 \textwidth]{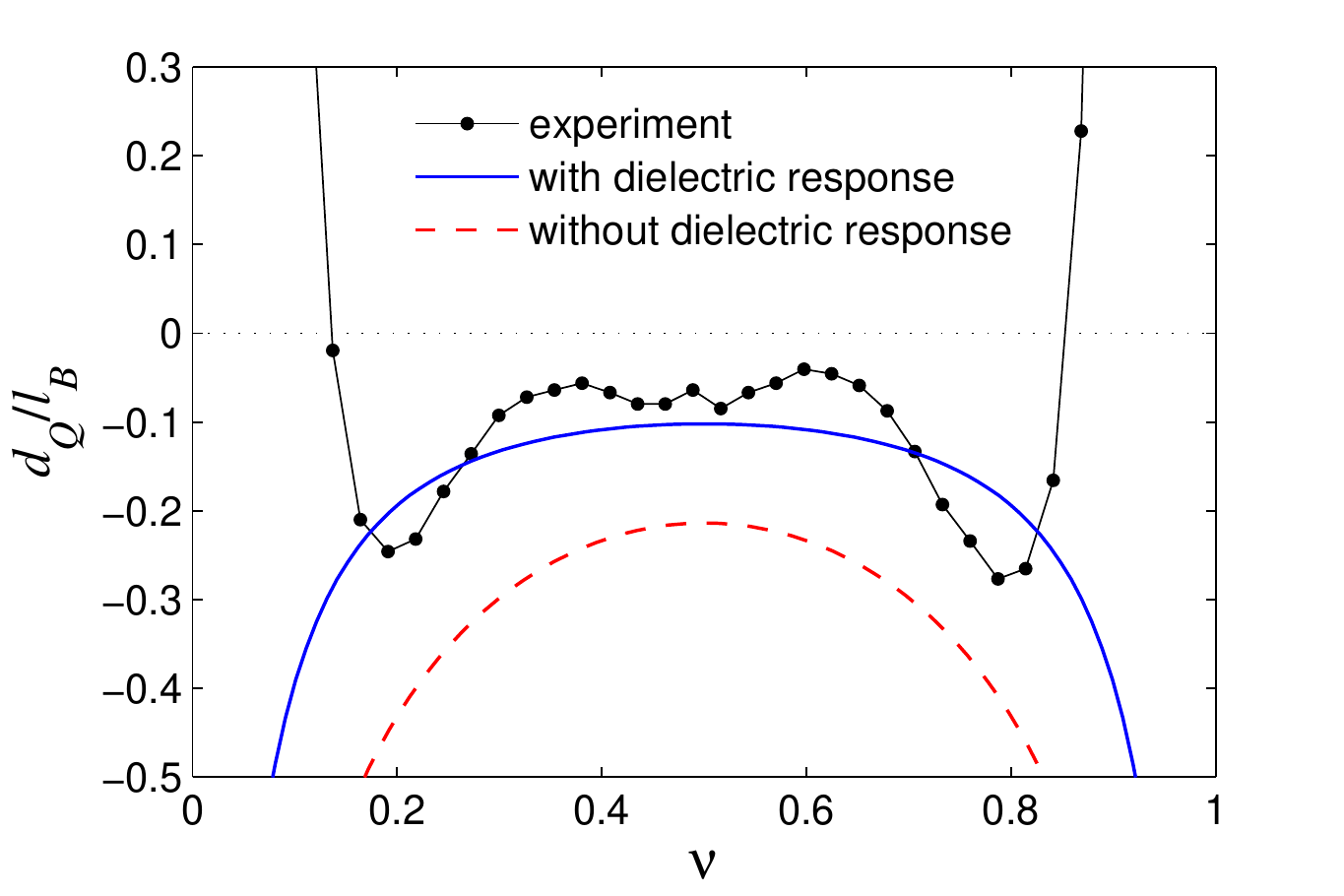}
\caption{(Color online) The quantum capacitance length $d_Q$, in units of magnetic length, as a function of filling factor of one of the LLLSs at $B = 17$ T.  The solid (blue) curve contains no free parameters.}
\label{fig:B17}
\end{figure}

On the other hand, when a large magnetic field is applied, $d_Q$ becomes negative, signaling the onset of strong positional correlations.  As an example, Fig.\ \ref{fig:B17} shows experimental measurements corresponding to $B = 17$ T, where the magnetic length $\lb = 6.2$ nm.  At such large fields the lowest Landau level (LLL) is split into four LLLSs, and in each of them $d_Q$ is negative throughout most of the LLLS.  Fig.\ \ref{fig:B17} shows data corresponding to one of these LLLSs, which has total filling factor $0 < \ntot <1$.  As one can see, $d_Q$ acquires a value of $d_Q \sim -0.1 \lb$ at $\nu = 1/2$, and at $\nu \sim 0.2, 0.8$ it becomes as large as $d_Q \sim -0.25 \lb$.  On the other hand, at very small values of $\nu$ or $1 - \nu$, $d_Q$ becomes positive.  This reversal in the sign of $d_Q$ is related to disorder, which in the limit where either the electron or hole concentration is very small becomes larger than the interaction energy between electrons/holes.  As a result, pores open up in the 2DEG and electric field lines originating at the gate electrode leak through the 2DEG, and consequently $d_Q$ grows sharply \cite{Shklovskii1986soo, Fogler2004nsa}.  Similar behavior was also observed in semiconductor quantum wells \cite{Eisenstein1992nco, Eisenstein1994cot} and in more recent experiments with graphene \cite{Yu2013ipi, Feldman2012usf, Feldman2013fqh, Hunt2013mdf}.  

One can also notice that the curve $d_Q(\nu)$ shows bumps centered around $\nu = 1/3$ and $\nu = 2/3$.  These bumps presumably arise from the cusps in the energy $E(\nu)$ associated with FQH states, as mentioned in the previous section, and are again consistent with previous observations\cite{Eisenstein1992nco, Eisenstein1994cot, Yu2013ipi, Feldman2012usf, Feldman2013fqh, Hunt2013mdf}.  

In the center of the LLLS, where the sample disorder plays a relatively small role, the observed quantum capacitance length is close to the predicted theoretical curve, as derived in Sec.\ \ref{sec:theory}.  For comparison, we also plot the value of $d_Q$ that arises if one ignores the dielectric response of the graphene [setting $\epsilon(q) = 1$ in the calculations of Sec.\ \ref{sec:theory}, as was done in Ref.\ \onlinecite{Skinner2013gcp}].  As one can see, this curve overestimates the magnitude of the quantum capacitance effect in the middle of the LLLS by more than two times.  

It is worth noting that, in general, the correct result for the quantum capacitance cannot be arrived at simply by replacing the substrate dielectric constant $\kappa$ in the expression for $E(\nu)$ by a constant value $\kappa (1 + \pi \alpha/2) \approx \kappa + 3.5$ [as in Eq.\ (\ref{eq:epsconst})].  Indeed, the dielectric function becomes constant only in the limit $q \gg 1/\lb$ (short-distance interactions), while the typical distance between interacting electrons becomes of order $\lb$ only in the middle of the LLLS.  As a result, the quantum capacitance is sensitive to the changing role of dielectric screening as a function of $\nu$, with the reduction of $d_Q$ relative to the unscreened state being largest at the middle of the LLLS.  As one approaches either $\nu = 0$ or $\nu = 1$, where the compressibility is related to the repulsion between distant electrons or holes, the role of dielectric response disappears.  Indeed, the solid (blue) curve in Fig.\ \ref{fig:B17} reproduces the calculation without dielectric response (dashed, red curve) in the limit of asymptotically small $\nu$ or $1-\nu$.

It may seem surprising to note from Fig.\ \ref{fig:B17} that at $\nu = 1/2$ the dielectric response reduces $d_Q$ by more than two times, while the maximum value of the dielectric function is only $1 + \pi \alpha/2 \approx 2.0$.  However, this large reduction in the quantum capacitance effect can be viewed as a consequence of the non-monotonic change in the influence of $\epsilon(q)$ with changing $\nu$, as discussed above.  One can understand the large reduction in $d_Q$ at $\nu \sim 1/2$ by first noting that, since the effect of the dielectric response disappears at $\nu \ll 1$, the curve $E(\nu)$ must coincide at very small $\nu$ with the one corresponding to $\epsilon(q) = 1$ (see Fig.\ \ref{fig:newFO}).  On the other hand, the dielectric response greatly reduces the energy at $\nu = 1$.  As a consequence, the curve $E(\nu)$ becomes particularly flat around $\nu \sim 1/2$, which implies that $d_Q$ is greatly reduced in the middle of the LLLS. 

\begin{figure}[htb!]
\centering
\includegraphics[width=0.5\textwidth]{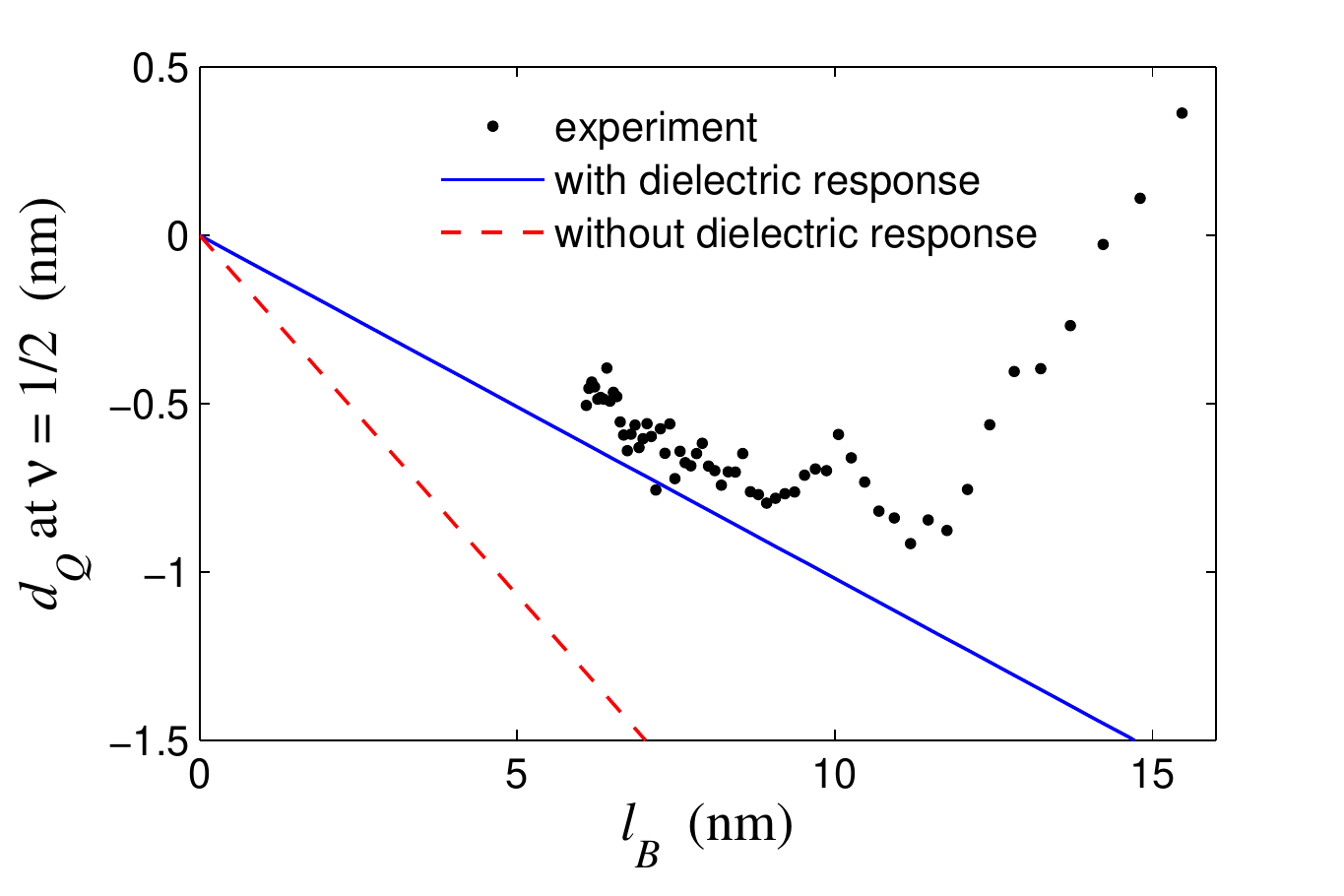}
\caption{(Color online) The quantum capacitance length $d_Q$ at $\nu = 1/2$ as a function of $\lb \propto B^{-1/2}$.  Experimental data is plotted only for the LLLS with $\nu = \ntot = 1/2$; the value of $d_Q$ for all four LLLSs is shown in Fig.\ \ref{fig:nu12-all}.  The solid (blue) line [Eq.\ (\ref{eq:dQ12})] contains no free parameters.}
\label{fig:nu12}
\end{figure}

We can also examine how the quantum capacitance in the middle of the LLLS depends on the magnetic field.  In Fig.\ \ref{fig:nu12} we plot the value of $d_Q$ at $\nu = 1/2$ for the LLLS with $\nu = \ntot$.  Our theoretical prediction, Eq.\ (\ref{eq:dQ12}), suggests that $d_Q$ should decline linearly with $\lb$ according to $d_Q \approx -0.10 \lb$ at $\nu = 1/2$.  And indeed, the experimental value of $d_Q$ seems to decline linearly with $\lb$ until $\lb \approx 12$\,nm ($B \approx 4.5$ T), at which point the magnetic field (and, correspondingly, the electron density) becomes small enough that electron correlations are washed out by disorder, and $d_Q$ rises abruptly.  

Experimentally, $\lb \approx 12$\,nm coincides approximately with the collapse of spin-valley splitting of the LLL.  As a consequence, at $\lb \gtrsim 12$\,nm the TDOS is essentially uniform across the LLL ($-2 < \ntot < 2$).  Nonetheless, there is a wide range of magnetic field, $0.25$\,T $< B < 4.5$\,T, in which the TDOS remains much larger than its $B = 0$ value, and correspondingly the $N = 0$ LL remains well separated from the adjacent $N = \pm 1$ levels.  As $B$ is decreased ($\lb$ increased) within this range, the TDOS decreases until it saturates at its $B = 0$ value, which corresponds to $d_Q \approx 17$\,nm and is determined by disorder-induced electron hole puddles \cite{Ponomarenko2010dos}, as mentioned above.  Experimentally, this saturation occurs at $\lb \approx 50$\,nm ($B \approx 0.25$\,T), and indeed coincides approximately with the collapse of Landau quantization.

The solid (blue) line in Fig.\ \ref{fig:nu12} shows the theoretical result of Eq.\ (\ref{eq:dQ12}), which is in close agreement with the experimental data in the regime $\lb \lesssim 12$\,nm.  As in Fig.\ \ref{fig:B17}, failing to account for the graphene dielectric response (red dashed line) leads to an overestimate of the quantum capacitance effect by more than two times.

\begin{figure}[htb!]
\centering
\includegraphics[width=0.48\textwidth]{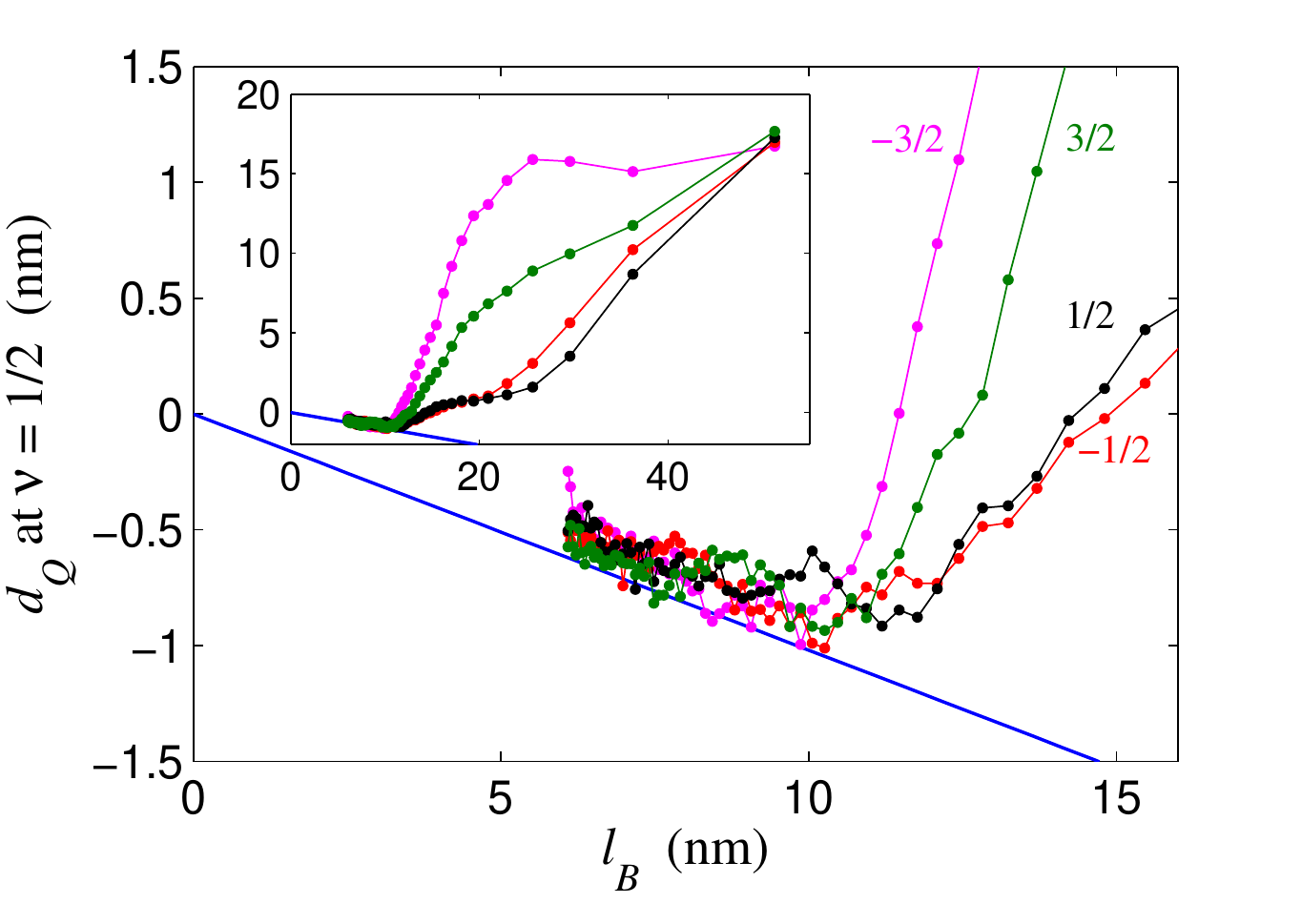}
\caption{(Color online) The quantum capacitance length $d_Q$ at $\nu = 1/2$ as a function of $\lb$ for all four LLLSs.  The straight (blue) line is the theoretical curve [Eq.\ (\ref{eq:dQ12})], and lines with points are experimental data.  These lines are labelled by their corresponding value of the total filling factor $\ntot$.  The inset shows the same data over a much wider range of $\lb$ (lower magnetic fields).}
\label{fig:nu12-all}
\end{figure}

Figs.\ \ref{fig:B17} and \ref{fig:nu12} show data from one of the four LLLSs, but one can also examine the quantum capacitance length for each of the other three.  As expected, we find that at large magnetic field $d_Q$ is essentially identical in all four LLLSs.  This equivalence is demonstrated in Fig.\ \ref{fig:nu12-all}, where $d_Q$ is plotted as a function of $\lb$ for total filling factors $\ntot = \pm 1/2, \pm 3/2$.  
One can see from this plot that at $\lb \lesssim 10$\,nm data from all four values of $\ntot$ collapse onto the same line as in Fig.\ \ref{fig:nu12}.

On the other hand, at larger $\lb$ (smaller magnetic feld),
where spin-valley splitting collapses and a single $N = 0$ LL emerges, one can see a big difference between $\ntot = \pm 3/2$ and $\ntot = \pm 1/2$.  In particular, at $10$\,nm $ < \lb < 20$\,nm the data corresponding to $\ntot = \pm 3/2$ show a much faster growth of $d_Q$ than the data for $\ntot = \pm 1/2$.  This difference can be understood qualitatively by thinking that once spin and valley splitting have collapsed, all four values of $\ntot$ correspond to the same LL, and that this LL has a bell-shaped density of states (DOS) as a function of energy.  Then at $\ntot = \pm 1/2$ the Fermi level is close to the LL center, thereby having larger DOS and smaller $d_Q$, while at $\ntot = \pm 3/2$ the Fermi level is in the tails of the DOS, and therefore has a smaller DOS and a larger $d_Q$.  When the magnetic field is made even smaller ($\lb < 20$\,nm), the $N = 0$ LL begins to merge with the adjacent $N = \pm 1$ LLs.  Eventually (at $\lb \sim 50$\,nm) this merging leads to a constant DOS, so that the curves for $d_Q(\lb)$ at $\ntot = \pm 1/2$ and $\pm 3/2$ merge.

\section{Discussion: Very thin and double-layer devices}
\label{sec:discussion}

In the preceding sections we presented results for $d_Q$ as a function of magnetic field and filling factor.  We reiterate here that our focus has been on the case of $d/\lb > 1$, and that the theory of Sec.\ \ref{sec:theory} loses its validity if its results are pushed to a regime where the typical distance between electrons, $n^{-1/2}$, is much larger than the distance $d$ to the metal gate.  In such a regime, the metal gate provides significant screening of the electron-electron interaction through the formation of image charges in the electrode surface \cite{Skinner2010alc, Skinner2010svm, Skinner2013gcp}.  As a result of this screening, 
the value of $d_Q$ in the LLLS saturates in the limit $\lb/d \gg 1$, so that $d + d_Q \propto (d^2/\lb)$ throughout the LLLS\cite{Skinner2013gcp}. 

The effect of the graphene dielectric response in the ``image charge regime" $d/\lb < 1$ remains to be explored, but in general its role should be weaker than for devices with large $d$.  Indeed, at $d/\lb \ll 1$ the extreme proximity of the metal gate means that the Coulomb interaction takes the form $\sim e^2 d^2/r^3$ at all distances $r \gg d$, and this weakened interaction should reduce the value of $\epsilon(q)$ at all $q \lesssim 1/d$, bringing it close to unity.  As a result, in the limit $d/\lb \ll 1$ the graphene dielectric response in the LLLS is essentially eliminated, and the results derived in Ref.\ \onlinecite{Skinner2013gcp} should be valid.

One can also consider how our results apply to capacitor devices made from two parallel graphene sheets with a voltage applied between them, rather than from a single graphene sheet with a metal gate electrode.  In the case where both graphenes are undoped (both have filling factor $\nu = 0$ in the absence of an applied voltage) and the distance $d$ between them is large enough that $\lb/d \ll 1$, one can effectively treat the two graphenes as independent 2DEGs.  In this case both layers contribute equally to the quantum capacitance, and one can arrive at the correct value of the quantum capacitance by simply doubling the value of $d_Q$ derived in Sec.\ \ref{sec:theory}.  For example, for a double-layer graphene device on hBN, Eq.\ (\ref{eq:dQ12}) suggests that $d_Q(\nu = 1/2) \approx -0.20 \lb$.  Of course, if one of the two graphenes is heavily doped (has a much larger carrier concentration than the other), then its TDOS becomes large and it essentially acts like a metal electrode, so that $d_Q$ is identical to that of a graphene-metal capacitor.

\begin{figure}[htb!]
\centering
\includegraphics[width=0.5\textwidth]{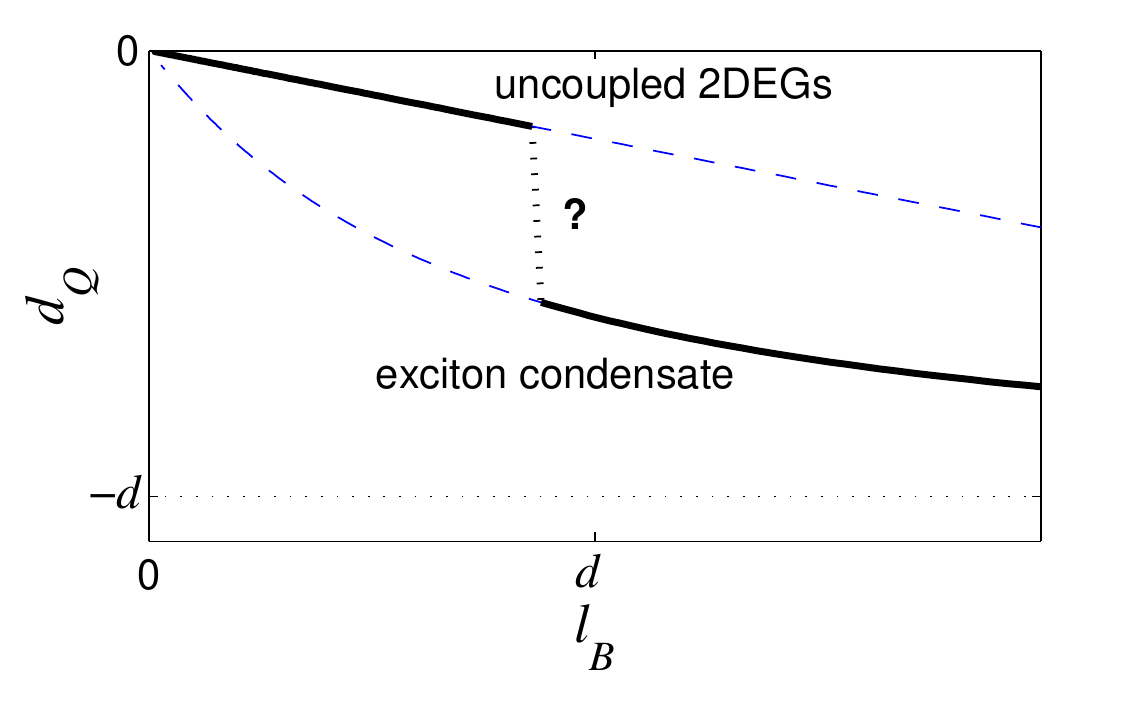}
\caption{(Color online) Schematic dependence of the quantum capacitance length $d_Q$ on the magnetic length $l_B$ for a capacitor made from two parallel graphene layers, plotted at fixed filling factor $\nu$ (for example, at $\nu = 1/2$).  At small $\lb/d$ (large magnetic field and thick devices), electrons in the two graphenes are essentially uncoupled, and their quantum capacitance can be captured by doubling the result for $d_Q$ given in Sec.\ \ref{sec:theory}.  At large $\lb/d$ (small magnetic field and thin devices) electrons in holes on opposite layers couple to each other, forming an exciton condensate.  This leads to stronger mutual screening and therefore\cite{Skinner2013gcp} to a larger negative $d_Q$.}
\label{fig:double-schematic}
\end{figure}

On the other hand, if $\lb/d$ is made large, then the system undergoes a phase transition to an exciton condensate state, in which electrons in one layer couple with holes in the opposite layer to form indirect excitons \cite{Lozovik1975fsp, Yoshioka1990dqw, Eisenstein2004bec}.  In this case, the interaction between neighboring excitons also becomes dipole-like, leading to an enhanced quantum capacitance (larger negative $d_Q$).  The critical value of $\lb/d$ for which exciton condensation first appears has been estimated to be slightly smaller than unity \cite{Yoshioka1990dqw, Joglekar2002bvi, Zhang2007sie}.  Using a naive description of the two phases, one can expect an abrupt change in $d_Q$ at the transition point, as illustrated schematically in Fig.\ \ref{fig:double-schematic}.  As of our writing, the details of this transition remain incompletely understood.  At very large $\lb/d$, the quantum capacitance length approaches $-d$ such that $d_Q +d \simeq 0.6 d^2/\lb$, similar to the case of thin graphene-metal devices discussed above \cite{Skinner2013gcp}.  As for single-layer graphene devices, the dielectric response can be expected to play a significantly weaker role in this limit.

Of course, experimental exploration of the small $d/\lb$ limit and the exciton condensate phase imposes a more stringent condition on the sample disorder.  For such thin devices the disorder competes not with the unscreened Coulomb interaction, of order $e^2/\lb$, but with the much weaker dipole interaction, $\sim e^2 d^2/\lb^3$.  Still, the exciton condensate phase may be first achieved already at $d/\lb \sim 1$.  In this regime, observation of enhanced negative $d_Q$ may be an early sign of the appearance of the exciton condensate phase.  (For hBN substrates, this is $d_Q < -0.20 \lb$ at $\nu = 1/2$.)

\section{Conclusion}
\label{sec:conclusion}

In this paper we have shown that the graphene dielectric response plays a prominent role in determining the quantum capacitance in a magnetic field.  In Sec.\ \ref{sec:theory} we suggested a simple, approximate method for taking this dielectric response into account, and in Secs.\ \ref{sec:experiment} and \ref{sec:compare} we presented experimental measurements of the quantum capacitance of graphene on boron nitride that support our theoretical results.

In the future, it would be worthwhile to study the quantum capacitance of graphene with different substrates, or in other words for a range of $\alpha$.  The role of dielectric response should be particularly large for suspended graphene, where $\alpha \approx 2.2$.  Data presented in Refs.\ \onlinecite{Feldman2012usf, Feldman2013fqh} suggest that for such experiments the value of $d_Q$ is indeed much closer to zero than would be predicted by a naive calculation of the energy neglecting dielectric response.  Still, a more careful analysis of this situation is warranted.  We caution that our interpolation procedure will likely have a larger numerical uncertainty for such large $\alpha$, since in such cases the dielectric response plays a larger role at intermediate $\nu$, which are described only indirectly.

Finally, we close with a comment on the effect of the dielectric response for the energy gap of the FQH states, focusing our discussion on the $\nu = 1/3$ state.  Theoretical predictions \cite{Toke2006fqh, Apalkov2006fqh, Shibata2008cca} generally estimate this energy gap to be in the range $\D = (0.03 - 0.1)(e^2/\kappa \lb)$.  On the other hand, experimental measurements of $\D$ for suspended graphene (where $\kappa = 1$) have reported\cite{Feldman2012usf, Ghahari2011m$f, Bolotin2009ofq, Du2009fqh} a significantly smaller value $\D = (0.008 - 0.02)(e^2/\lb)$.  (For a summary of the comparison between theory and experiment, see the supplemental material of Ref.\ \onlinecite{Feldman2012usf}.)  One likely source of this discrepancy is the graphene dielectric response, which reduces the strength of interactions at large distances $r \gtrsim \lb$, and therefore tends to reduce $\D$.  While we have not made an attempt to calculate the value of $\D$, one can obtain a rough estimate by conjecturing that $\D$ should be proportional to the total energy $E(\nu=1/3)$.  Using the methods of Sec.\ \ref{sec:theory}, we find that for suspended graphene the dielectric response reduces the value of $E(1/3)$ by a factor of $\sim 2.2$ (from $-0.41 e^2/\lb$ to $-0.19 e^2/\lb$).  Thus, one can arrive at a crude estimate for $\D$ by dividing the above theoretical prediction range by $2.2$, which gives
\be 
\D = (0.015 - 0.045)e^2/\lb.
\ee 
This revised estimate is closer in line with experiment.
We emphasize again that this factor $2.2$ is different from what one would get by replacing the vacuum dielectric constant $\kappa = 1$ with the constant factor $1 + \pi \alpha/2 \approx 4.5$.  The true effect of the dielectric response in a magnetic field is somewhat smaller, as resulting from a weighted integration of $\epsilon(q)$ over all wave vectors.

\acknowledgments

We are grateful to 
B. Feldman,
M. M. Fogler,
A. M. Straub,
and A. Yacoby
for helpful discussions.
This work was supported primarily by the National Science Foundation through the University of Minnesota MRSEC under Award Number DMR-0819885.  B.S. and B.I.S. thank the Aspen Center for Physics for their hospitality, and acknowledge their NSF grant 1066293.

\appendix

\section{Parameterized values of the coefficients of the Fano-Ortolani expression}
\label{app:FO}

In Sec.\ \ref{sec:theory} we explained our numerical procedure for calculating the energy $E(\nu)$ for a 2DEG in the LLLS.  Here we list our calculated numerical values of the coefficients that enter the FO expression, Eq.\ (\ref{eq:FO}).  The formulas presented below correspond to parameterizations of numerical calculations performed over the range $0 \leq \alpha \leq 2.2$.

The energy of the $\nu = 1$ state, calculated according to Eq.\ (\ref{eq:E1}), is given approximately by
\be
\frac{E(1)}{e^2/\kappa \lb} \approx - \sqrt{\frac{\pi}{8}} \exp \left[ -0.927 \alpha + 0.379 \alpha^2 - 0.0751 \alpha^3 \right].
\ee
The coefficients $a_3$, $a_4$, and $a_5$ can be parameterized as
\begin{eqnarray}
a_3 & = & -0.773 - (2.01 \times 10^{-3}) \alpha + (4.74 \times 10^{-3}) \alpha^2 \nonumber \\
& &  + (6.60 \times 10^{-4}) \alpha^3 \\
a_4 & = & 0.460 + 0.412 \alpha + 0.173 \alpha^2 - 0.0687 \alpha^3 \\
a_5 & = & -0.198 + 0.193 \alpha - 0.742 \alpha^2 + 0.195 \alpha^3.
\end{eqnarray}
Using the equations above one can accurately reproduce our numerically calculated values of the energy $E(\nu)$ to within $1\%$ for all $\nu$ and all $\alpha \leq 2.2$.

\section{Energy of the Wigner crystal state}
\label{app:hartree}

We are interested in describing the energy of the Wigner crystal state at small filling factors, $\nu < \nuc = 0.2$.  At such small $\nu$, the exchange interaction $E_{ex}$ is exponentially small (in the absence of dielectric screening, $E_{ex}/(e^2/\kappa \lb) = -1.4 \exp[-2.07/\nu]$)\cite{Lam1984lta}, and can therefore safely be ignored.  The total energy $E_{WC}$ can thus be approximated to high accuracy using just the semiclassical (Hartree) approximation.

In the Wigner crystal state, the electron wavefunctions $\varphi_{\RR}(\rr)$ can be described as Gaussian wave packets centered at the points of the (triangular) Wigner lattice:
\be 
\varphi_{\RR}(\rr) = \frac{1}{\sqrt{2 \pi \lb^2}} \exp \left[ - \frac{|\rr - \RR|^2}{4 \lb^2} \right].
\ee
Here, $\RR$ is a vector indicating one of the lattice points.  The Hartree energy can be written as
\be 
E_{WC} = \frac12 \int d^2 \rr d^2 \rr' V(\rr - \rr') \left| \varphi_{\bf{0}}(\rr) \right|^2 \sum_{\RR \neq \mathbf{0}} \left| \varphi_{\RR}(\rr') \right|^2 ,
\ee 
where $\varphi_{\bf{0}}(\rr)$ denotes the wavefunction of the electron at the origin.

Writing $V(r)$ in terms of its Fourier transform and evaluating the sum gives
\be 
E_{WC} = \frac{n}{2} \sum_{q \in G} \Vt(q) e^{-q^2 \lb^2} - \frac12 \int_0^\infty \frac{q \Vt(q)}{2 \pi} e^{-q^2\lb^2} dq,
\label{eq:EH}
\ee
where $G$ denotes the set of all nonzero reciprocal lattice vectors of the triangular Wigner lattice.  The second term on the right-hand side of Eq.\ (\ref{eq:EH}) comes from removing the self-interaction term from the Hartree energy ($\RR = \mathbf{0}$).  Substituting $\Vt(q) = 2 \pi e^2/[ \kappa \epsilon(q) q ]$ and $n = \nu/2 \pi \lb^2$ into Eq.\ (\ref{eq:EH}) gives
\begin{eqnarray} 
\frac{E_{WC}(\nu)}{e^2/\kappa \lb} & = & \frac{\nu}{2} \sum_{q \in G} \frac{ \exp [- (q \lb)^2 ]}{q \lb \epsilon(q)} \label{eq:EHdetail} \\ 
& & - \frac12 \int_0^\infty \frac{\exp[-(k \lb)^2/2]}{\epsilon(k)} \lb dk. \nonumber
\end{eqnarray} 
The reciprocal lattice vectors $q \in G$ can be labeled with integer indices $i,j$ such that the wave vectors $q_{ij}$ in the sum are
\be 
q_{ij} = \frac{1}{\lb} \sqrt{ \frac{4 \pi \nu}{\sqrt{3}} (i^2 + i j + j^2) },
\ee
so that the sum in Eq.\ (\ref{eq:EHdetail}) is over all $\{i,j\} \neq \{0, 0\}$.

Eq.\ (\ref{eq:EHdetail}) is evaluated numerically for a range of $\nu$ corresponding to $0 < \nu < \nuc $, as discussed in Sec.\ \ref{sec:theory}.  An example calculation is shown as the thick (blue) line in Fig.\ \ref{fig:newFO}.

\section{Calculation of the energy of the $\nu = 1/5$ and $\nu = 1/3$ fractional quantum Hall states}
\label{app:fqh}

The general expression for the interaction energy per electron is given in Eq.\ (\ref{eq:Eg}), which can be rewritten as
\be 
E(\nu)  = \frac{\nu}{4 \pi \lb^2} \int d^2r V(r) [g_{\nu}(r) - 1],
\label{Eq:Egnu}
\ee
where $g_{\nu}(r)$ is the pair distribution function for a given filling factor.  Writing the interaction law $V(r)$ in terms of its Fourier transform gives
\be 
V(r) = \int \frac{d^2 q}{(2\pi)^2} \Vt(q) \exp[i {\bf q} \cdot {\bf r}] = \frac{e^2}{\kappa} \int_0^\infty dq \frac{J_0(q r)}{\epsilon(q)},
\ee 
where $J_0(x)$ is the zeroth order Bessel function of the first kind.  Inserting this expression into Eq.\ (\ref{Eq:Egnu}), one arrives at a general expression for the energy:
\be 
E(\nu) = \frac{\nu e^2}{2 \kappa \lb^2} \int_0^\infty dq \int_0^\infty dr \frac{r [g_\nu(r) - 1] J_0(q r)}{\epsilon(q)}.
\label{eq:EFQH}
\ee 

For the FQH states at $\nu = 1/5$ and $\nu = 1/3$, the pair distribution functions, $g_{1/5}(r)$ and $g_{1/3}(r)$, respectively, have been parameterized from Monte Carlo data \cite{Girvin1986mrt}.  Using this parameterized result allows us  to evaluate Eq.\ (\ref{eq:EFQH}) numerically.  As shown in Fig.\ \ref{fig:newFO}, the resulting energies align fairly closely with the result of our interpolation method.

A similar calculation was performed in Ref.\ \onlinecite{Fogler1997llt}, where the effect of $\epsilon(q)$ was incorporated into calculations of the energy of conventional semiconductor 2DEGs.

\bibliography{qc}

\end{document}